\documentstyle[prd,aps]{revtex}

\title{A New Expansion of the Heisenberg Equation of Motion \\
with Projection Operator}

\author{T. Koide and M. Maruyama}

\begin{document}

\maketitle

\begin{abstract}
We derive a new expansion of the Heisenberg equation of motion based 
on the projection operator method proposed by Shibata, Hashitsume and Shing\=u.
In their projection operator method, 
a certain restriction is imposed on 
the initial state.
As a result, one cannot prepare arbitrary initial states, for example a coherent state,  
to calculate the time development of quantum systems.
In this paper, we generalize the projection operator method 
by relaxing this restriction.
We explain our method in the case of a Hamiltonian both with and without explicit 
time dependence.
Furthermore, we apply it to an exactly solvable model called 
the damped harmonic oscillator model and confirm the validity of our method. 
\end{abstract}

\section{Introduction}

It is important 
to calculate the time evolution in quantum field theory 
in various fields of physics.
However, there is no established method to carry out such calculations.
In this work, we study a new expansion of the Heisenberg equation of motion 
based on the projection operator method.
The projection operator method, which we use in this paper, 
was first proposed by Shibata, Hashitsume and Shing\=u. \cite{ref:1}

The projection operator method has several important characteristics.
(i) It can be applied both in Schr\"odinger and Heisenberg pictures.
(ii) There is an arbitrariness in the selection of projection operators.
(iii) Equations with and without time convolution integrals are treated systematically.
(iv) One can use it to calculate in the case of an explicitly time-dependent Hamiltonian.
The method has been successfully applied to problems 
in quantum mechanics. \cite{ref:2}
The Mori and Nakajima-Zwanzig methods,\cite{ref:10} 
which are well-known methods using the projection operator, 
are included in this projection operator method.
We believe that the projection operator method is 
more convenient than other common formalisms, e.g., 
the closed time-path formalism\cite{ref:3} and the 
Feynman-Vernon influence functional technique.\cite{ref:4}
In such formalisms, further approximation is needed 
to derive an equation of motion 
without a time convolution integral, for example, 
as performed in the physics of hadrons and the early universe.\cite{ref:5}
However, in the projection operator method, there is no such problem, 
because of the above characteristic (iii).
Furthermore, since this method consists of canonical formalism, 
it is useful to discuss the meanings of time evolution 
in quantum field theory by comparing with usual scattering theory.

In a previous paper, 
we discussed the renormalization of ultraviolet divergences 
in a certain model, using an equation of motion without a time convolution 
integral derived with the projection operator method.\cite{ref:6}
However, there is still the problem that 
we cannot prepare arbitrary initial states, 
because a restriction is 
imposed on the initial state 
to carry out the systematic expansion of the Heisenberg equation of motion.
In this work, we study a new expansion of the Heisenberg equation of motion 
based on the projection operator method without imposing the restriction.
For this reason, we can prepare more general initial states, such as coherent states, 
which may be important to describe phase transitions.

This paper is organized as follows.
In \S $2$, we explain the formalism in the case that Hamiltonian has 
no explicit time dependence.
In \S $3$, the case in which Hamiltonian has explicit time dependence 
is discussed.
Such a case is also important for studying physical phenomena, for example, 
pair creation in a strong external field.
%The derivation in this case is already discussed in ref \cite{ref:9}.
In \S $4$, we apply our formalism to the damped harmonic oscillator model, 
which is exactly solvable, to confirm the validity of our method.
Conclusions are given in \S $5$.

\section{The Heisenberg equation with time-independent Hamiltonian}

Our strategy is to study quantum field theory as an initial value problem.
In this section, we derive our equation in the case that the Hamiltonian 
has no explicit time dependence.
Our starting point is the Heisenberg equation of motion,
\begin{eqnarray}
  \frac{d}{dt}O(t) &=& i[H,O(t)] \\
                   &=& iLO(t) \\
   \longrightarrow O(t) &=& e^{iL(t-t_{0})}O(t_{0}), \label{eqn:HE}
\end{eqnarray}
where $L$ is the Liouville operator and $t_{0}$ is the time at which 
we prepare an initial state.
The Heisenberg equation contains complete information of the time 
evolution of the operator, but in general, 
it is difficult to solve exactly when there are interactions.
Therefore it is necessary to make some approximations.
For this purpose, we introduce generic projection operators $P$ and $Q$ 
that have the following general properties:
\begin{eqnarray}
   P^2 &=& P, \\
   Q &=& 1-P, \\
   PQ &=& QP = 0.
\end{eqnarray}
By using these projection operators, we can carry out coarse-grainings 
in the time development.
From Eq. (\ref{eqn:HE}), one can see that the time dependence of the operators 
is determined by $e^{iL(t-t_{0})}$.
This yields the equation
\begin{eqnarray}
   \frac{d}{dt}e^{iL(t-t_{0})} &=& e^{iL(t-t_{0})}iL \nonumber \\
                &=& e^{iL(t-t_{0})}(P+Q)iL.   \label{eqn:P+Q}
\end{eqnarray}
From this equation, we can derive the following two equations:
\begin{eqnarray}
  \frac{d}{dt}e^{iL(t-t_{0})}P 
&=& e^{iL(t-t_{0})}PiLP +e^{iL(t-t_{0})}QiLP, \label{eqn:P}\\
  \frac{d}{dt}e^{iL(t-t_{0})}Q 
&=& e^{iL(t-t_{0})}PiLQ +e^{iL(t-t_{0})}QiLQ. \label{eqn:Q}
\end{eqnarray}
Equation (\ref{eqn:Q}) can be solved for $e^{iL(t-t_{0})}Q$.
We obtain
\begin{eqnarray}
\hspace*{-2cm}  e^{iL(t-t_{0})}Q &=& Qe^{iLQ(t-t_{0})} + \int^{t}_{t_{0}}ds e^{iL(s-t_{0})}PiLQe^{iLQ(t-s)} \label{eqn:TC-dainyuu}\\
           &=& Qe^{iLQ(t-t_{0})} + e^{iL(t-t_{0})}(P+Q)\int^{t}_{t_{0}}ds e^{-iL(t-s)}PiLQe^{iLQ(t-s)} \nonumber \\
           &=& Qe^{iLQ(t-t_{0})}\frac{1}{1-\Sigma(t,t_{0})}+e^{iL(t-t_{0})}P\Sigma(t,t_{0})\frac{1}{1-\Sigma(t,t_{0})},\label{eqn:TCL} 
\end{eqnarray}
where 
\begin{eqnarray}
     \Sigma(t,t_{0}) &\equiv& \int^{t}_{t_{0}}d{s}e^{-iL(t-s)}PiLQe^{iLQ(t-s)}.
\end{eqnarray}
Substituting Eq. (\ref{eqn:TCL}) into Eq. (\ref{eqn:P+Q}) and operating with $O(t_{0})$ from the right, 
we obtain 
\begin{eqnarray}
     \frac{d}{dt}O(t) 
&=& e^{iL(t-t_{0})}PiLO(t_{0})+e^{iL(t-t_{0})}P\Sigma(t,t_{0})\frac{1}{1-\Sigma(t,t_{0})}iLO(t_{0}) \nonumber \\
&&+ Qe^{iLQ(t-t_{0})}\frac{1}{1-\Sigma(t,t_{0})}iLO(t_{0}). \nonumber \\
\label{eqn:genmitu-nt}
\end{eqnarray}
This equation is equivalent to the Heisenberg equation, 
and it has no time convolution integral.
We demonstrate this point at the end of this section.
Therefore, we call this the time-convolutionless (TCL) equation.
It is the same equation as that used 
in the previous projection operator method.\cite{ref:1}

Now we consider the situation that the total system can be divided into 
two parts: the system and the environment.
We want to know the detailed behavior of the degrees of freedom of the system.
Therefore, we carry out coarse-grainings to treat 
the environment degrees of freedom, as in the Feynman-Vernon influence 
functional technique \cite{ref:4} 
and the projected effective Hamiltonian. \cite{ref:7}
Our next task is to specify the projection operator for this purpose.
In this case, the total Hamiltonian can be divided into three parts, 
the system (S), the environment (E) and the interaction (I) between 
the system and the environment:
\begin{eqnarray}
  H = H_{S} + H_{E} + H_{I}.
\end{eqnarray}
The self-interaction of the system and/or the environment 
can also be included in $H_{I}$.
Furthermore, we assume that the initial density matrix $\rho$ is given by 
the direct product of the system density matrix $\rho_{S}$ 
and the environment density matrix $\rho_{E}$, 
all given at the initial time $t_{0}$:
\begin{eqnarray}
  \rho = \rho_{S}\otimes\rho_{E}. \label{eqn:ini-density}
\end{eqnarray}
We then define the projection operator as 
\begin{eqnarray}
  PO = {\rm Tr}_{E}[\rho_{E}O] \equiv \langle O \rangle_{E} 
\label{eqn:ex-pro}
\end{eqnarray}
for any operator $O$.
With this projection, we can replace an operator acting on 
the environment with a c-number.
From the nature of the projection operator, 
we obtain the following relations:
\begin{eqnarray}
  PL_{S} &=& L_{S}P, ~~~~~~QL_{S} = L_{S}Q, \\
  L_{E}P &=& 0, ~~~~~~~~~~L_{E}Q = L_{E}, 
\end{eqnarray}
where $L_{a} O = [H_{a},O]$ for $a = S,E,I$.
Using these properties of the projection operator, 
we have the following relations:
\begin{eqnarray}
  QL_{0}Q &=& QL_{0}, \label{eqn:nrel} \\
  Qe^{iL_{0}t}Q &=& Qe^{iQL_{0}Qt}Q = Qe^{iL_{0}t},\label{eqn:nreld}
\end{eqnarray}
where $L_{0} = L_{S} + L_{E}$.

In the previous projection operator method, 
the additional relation $PL_{E}=0$ is assumed 
to derive a systematic expansion.\cite{ref:1}
This relation is satisfied only in restricted cases, 
for example, 
when $\rho_{E}$ is an eigenstate of $H_{E}$ or a mixed state which includes 
diagonal components in the basis of eigenstates of $H_{E}$. 
Therefore, this relation is not satisfied for general $\rho_{E}$.
Now, we take a coherent state for $\rho_{E}$ as an example:
\begin{eqnarray}
\rho_{E}=e^{-|\alpha|^2}
e^{\alpha a^{\dagger}}
|0\rangle \langle 0|e^{\alpha^{*}a}
\equiv|\alpha\rangle \langle\alpha|.
\end{eqnarray}
The state $|\alpha\rangle$ is an eigenstate of the 
annihilation operator $a$:
\begin{eqnarray}
 a|\alpha\rangle = \alpha|\alpha\rangle.
\end{eqnarray}
The quantity $PL_{E}O$ is calculated as 
\begin{eqnarray}
  PL_{E}O &=& {\rm Tr}_{E}[\rho_{E}L_{E}O] \nonumber \\
          &=& \langle\alpha |[H_{E}O-OH_{E}]|\alpha \rangle \nonumber \\
          &=& E\alpha^* 
\langle\alpha |a O|\alpha \rangle 
- E \alpha
\langle\alpha |O a^{\dagger}|\alpha \rangle, 
\label{eqn:coh-ex}
\end{eqnarray}
where we assume $H_{E} = Ea^{\dagger}a$.
The r.h.s.~of the above equation 
does not become zero for an arbitrary operator $O$.
Therefore, we can see that the relation $PL_{E}=0$ is not satisfied.
From these facts, we can conclude that the expansion used 
in the previous projection operator method is not applicable to 
general initial states.
We will derive a systematic expansion of the Heisenberg equation of motion 
without imposing the relation $PL_{E}=0$.
With this, we can calculate the time evolution by using 
more general initial states, but with 
the assumption Eq. (\ref{eqn:ini-density}).

With the above properties of the projection operators 
(\ref{eqn:nrel}) and (\ref{eqn:nreld}), 
the function $\Sigma(t,t_{0})$ can be expressed as 
\begin{eqnarray}
  \Sigma(t,t_{0}) 
       &=& \int^{t}_{t_{0}}d{s}e^{-iL(t-s)}PiLQe^{iLQ(t-s)} \nonumber \\
       &=& Q-e^{-iL(t-t_{0})}Qe^{iLQ(t-t_{0})} \nonumber \\
       &=& Q - e^{-iL_{0}(t-t_{0})}{\cal C}(t,t_{0})Q{\cal D}(t,t_{0})
e^{iL_{0}(t-t_{0})} \nonumber \\
       &=& Q - e^{-iL_{0}(t-t_{0})}Qe^{iL_{0}(t-t_{0})} 
- e^{-iL_{0}(t-t_{0})}({\cal C}(t,t_{0})-1)Q
e^{iL_{0}(t-t_{0})} \nonumber \\
       &&- e^{-iL_{0}(t-t_{0})}
Q({\cal D}(t,t_{0})-1)e^{iL_{0}(t-t_{0})} \nonumber \\
       && - e^{-iL_{0}(t-t_{0})}({\cal C}(t,t_{0})-1)Q({\cal D}(t,t_{0})-1)
e^{iL_{0}(t-t_{0})}.\label{eqn:sigma}
\end{eqnarray}
Here, the functions ${\cal C}(t,t_{0})$ and ${\cal D}(t,t_{0})$ are expressed 
as 
\begin{eqnarray}
  {\cal C}(t,t_{0}) 
&=& e^{iL_{0}(t-t_{0})}e^{-iL(t-t_{0})} \nonumber \\
&=& 1+\sum^{\infty}_{n=1}(-i)^n \int^{t}_{t_{0}}dt_{1}\int^{t_{1}}_{t_{0}}dt_{2} \cdots \int^{t_{n-1}}_{t_{0}}dt_{n} \breve{L}_{I}(t_{1}-t_{0})\breve{L}_{I}(t_{2}-t_{0}) \cdots\nonumber \\
&&\times  \breve{L}_{I}(t_{n}-t_{0}), \nonumber \\
\\
   {\cal D}(t,t_{0}) 
&=& e^{iQLQ(t-t_{0})}e^{-iQL_{0}Q(t-t_{0})} \nonumber \\
&=& 1+\sum^{\infty}_{n=1} i^n \int^{t}_{t_{0}}dt_{1}\int^{t_{1}}_{t_{0}}dt_{2} \cdots \int^{t_{n-1}}_{t_{0}}dt_{n} Q\breve{L}^{Q}_{I}(t_{n}-t_{0})Q\breve{L}^{Q}_{I}(t_{n-1}-t_{0}) \cdots \nonumber \\
&&\times Q\breve{L}^{Q}_{I}(t_{1}-t_{0}), \nonumber \\ 
\end{eqnarray}
where
\begin{eqnarray}
    \breve{L}_{I}(t-t_{0}) &\equiv& e^{iL_{0}(t-t_{0})}L_{I}e^{-iL_{0}(t-t_{0})}, \label{eqn:interaction}\\
    \breve{L}^{Q}_{I}(t-t_{0}) &\equiv& e^{iL_{0}(t-t_{0})}L_{I}Qe^{-iL_{0}(t-t_{0})}.
\end{eqnarray}
The operator ${\cal C}(t,t_{0})$ [${\cal D}(t,t_{0})$] is 
a time [an anti-time] ordered function of Liouville operators.
These expressions are derived in Appendix A.
Then, we have 
\begin{eqnarray}
     P\Sigma(t,t_{0})\frac{1}{1-\Sigma(t,t_{0})} 
&=& P\Sigma(t,t_{0})\frac{1}{1-Q\Sigma(t,t_{0})} \nonumber \\
&=& -Pe^{-iL_{0}(t-t_{0})}{\cal C}(t,t_{0})Qe^{iL_{0}(t-t_{0})} \nonumber \\
&&\times     \frac{1}{1+e^{-iL_{0}(t-t_{0})}({\cal C}(t,t_{0})-1)Qe^{iL_{0}(t-t_{0})}}. \nonumber \\
\label{eqn:Psigma}
\end{eqnarray}
Note that the second line on the r.h.s.~does not include ${\cal D}(t,t_{0})$.
The detailed derivation of Eq. (\ref{eqn:Psigma}) appears in Appendix B.
Substituting Eq. (\ref{eqn:Psigma}) into Eq. (\ref{eqn:genmitu-nt}), 
we obtain the final expression of the Heisenberg equation:
\begin{eqnarray}
 \frac{d}{dt}O(t) 
&=&  e^{iL(t-t_{0})}PiLO(t_{0}) \nonumber \\
&&  -e^{iL(t-t_{0})}Pe^{-iL_{0}(t-t_{0})}{\cal C}(t,t_{0})Q
     \frac{1}{1+({\cal C}(t,t_{0})-1)Q}
     e^{iL_{0}(t-t_{0})}iLO(t_{0}) \nonumber \\
&&  + Qe^{iLQ(t-t_{0})}\frac{1}{1-\Sigma(t,t_{0})}iLO(t_{0}).
\label{eqn:genmitu-nt-2}
\end{eqnarray}

When we expand $P\Sigma(t,t_{0})/(1-\Sigma(t,t_{0}))$ 
up to first order in the interaction $H_{I}$, 
we have
\begin{eqnarray}
  \frac{d}{dt}O(t) &=& e^{iL(t-t_{0})}Pe^{-iL_{0}(t-t_{0})}Pe^{iL_{0}(t-t_{0})}iLO(t_{0}) \nonumber \\
   && + e^{iL(t-t_{0})}Pe^{-iL_{0}(t-t_{0})}P\int^{t}_{t_{0}}ds e^{iL_{0}(s-t_{0})}iL_{I}e^{-iL_{0}(s-t_{0})}Qe^{iL_{0}(t-t_{0})}iLO(t_{0}) \nonumber \\
   && + Qe^{iLQ(t-t_{0})}\frac{1}{1-\Sigma(t,t_{0})}iLO(t_{0}). 
\label{eqn:t-i-res}
\end{eqnarray}
Here, we do not expand the third term on the r.h.s.~of Eq. (\ref{eqn:t-i-res}).
This term becomes zero when we take the expectation value, 
and therefore we do not expand it.
This expanded equation is used in \S 4.

It can be seen that Eq. (\ref{eqn:t-i-res}) does not contain 
a time convolution integral, 
because of the form of the full time-evolution operator, $e^{iL(t-t_{0})}$, 
which operates 
from the left in the second term on the r.h.s.~of the equation.
If this did contain a time convolution integral, 
the form of the full time-evolution operator must be $e^{iL(t-s)}$, where 
$s$ is an integral variable.
Such a time-convolution (TC) equation is discussed in Appendix C. 
In other formulations, e.g., 
the closed time-path formalism and the Feynman-Vernon influence functional technique, 
the derived equation of motion has a time convolution integral in general.
Such a time convolution term is called a ``memory term''.
The derived equation of motion is often solved using 
the Markov approximation.\cite{ref:5}
However, in our improved projection operator method 
(and the previous projection operator method), 
the equation without a time convolution integral, that is, the TCL equation 
is automatically obtained and it is not necessary to make the Markov approximation.

We would like to make some remarks regarding the operator $O(t_{0})$.
First, the choice of the operator $O(t_{0})$ 
is not restricted to an operator of the system.
We can choose an environment operator or a product of system 
and environment operators as the operator $O(t_{0})$.
This is different from the case of the previous projection operator method and 
the Uchiyama-Shibata (U-S) projection operator method, 
which was proposed recently,\cite{ref:9} 
because those method use the condition $PO(t_{0})=O(t_{0})$, which is satisfied 
only when $O(t_{0})$ is a system operator.
Therefore, in our improved projection operator method, 
it is possible to examine the time evolution of 
conserved quantities which are composed of not only system operators 
but also environment operators.
Any operator $O(t_{0})$ which commutes with the total Hamiltonian 
is conserved, for any order of expansion, 
because $LO(t_{0}) = 0$ in Eqs. (\ref{eqn:genmitu-nt}) and (\ref{eqn:genmitu-nt-2}).
This is a reasonable result, because conserved quantities should be 
time independent.

Until now, we have used the projection operator 
defined in Eq. (\ref{eqn:ex-pro}).
However, to derive Eq. (\ref{eqn:genmitu-nt-2}), 
only the condition (\ref{eqn:nrel}) is needed.
Therefore, we can use a more general projection operator to derive 
Eq. (\ref{eqn:genmitu-nt-2}) in the case that the condition (\ref{eqn:nrel}) 
is met.

\section{The Heisenberg equation for a time-dependent Hamiltonian}

In this section, we consider the case of a Hamiltonian with 
explicit time dependence.
We distinguish the time dependence of operators 
from explicit time dependence.
To do this, the time dependent Hamiltonian $H(t)$ is written with two time arguments: 
\begin{eqnarray}
H(t) = H(t,t),
\end{eqnarray}
where the first argument represents the explicit time dependence, 
and the second that of operators.

The time development of any operator $O$ which has no explicit time dependence 
can be expressed as
\begin{eqnarray}
  O(t) &=& e^{i\int^{t}_{t_{0}}ds L(s,t_{0})}_{\rightarrow} O(t_{0}), \label{eqn:ot}
\end{eqnarray}
where $t_{0}$ is the time at which we prepare an initial state.
The operator $e^{i\int^{t}_{t_{0}}d\tau L(\tau,t_{0})}_{\rightarrow}$ is defined as follows:
\begin{eqnarray}
   e^{i\int^{t}_{t_{0}}d\tau L(\tau,t_{0})}_{\rightarrow} &=& 
     1 + \sum^{\infty}_{n=1} i^n\int^{t}_{t_{0}}dt_{1}\int^{t_{1}}_{t_{0}}dt_{2}\cdots\int^{t_{n-1}}_{t_{0}}dt_{n}L(t_{n},t_{0})L(t_{n-1},t_{0})\cdots L(t_{1},t_{0}). \label{eqn:t-e} \nonumber \\
\end{eqnarray}
Similarly, $e^{i\int^{t}_{t_{0}}d\tau L(\tau,t_{0})}_{\leftarrow}$ 
is defined as 
\begin{eqnarray}
   e^{i\int^{t}_{t_{0}}d\tau L(\tau,t_{0})}_{\leftarrow} &=& 
     1 + \sum^{\infty}_{n=1} i^n\int^{t}_{t_{0}}dt_{1}\int^{t_{1}}_{t_{0}}dt_{2}\cdots\int^{t_{n-1}}_{t_{0}}dt_{n}L(t_{1},t_{0})L(t_{2},t_{0})\cdots L(t_{n},t_{0}). \nonumber \\
\end{eqnarray}
Therefore, the time evolution of the operator included 
in the Hamiltonian is expressed as 
\begin{eqnarray}
  H(t,t) &=& e^{i\int^{t}_{t_{0}}ds L(s,t_{0})}_{\rightarrow} H(t,t_{0}).
\end{eqnarray}
By using this relation, 
it becomes clear that the operator $O(t)$ given in Eq.~(\ref{eqn:ot}) 
satisfies the Heisenberg equation
\begin{eqnarray}
  \frac{d}{dt}O(t) 
&=& e^{i\int^{t}_{t_{0}}ds L(s,t_{0})}_{\rightarrow} iL(t,t_{0})O(t_{0}) \nonumber \\
&=& e^{i\int^{t}_{t_{0}}ds L(s,t_{0})}_{\rightarrow} i[H(t,t_{0}),O(t_{0})] \nonumber  \\
&=& i[H(t,t),O(t)].
\end{eqnarray}

Note the following properties:
\begin{eqnarray}
 \frac{d}{ds}e^{i\int^{t}_{s}d\tau L(\tau,t_{0})}_{\rightarrow} &=& -iL(s,t_{0})e^{i\int^{t}_{s}d\tau L(\tau,t_{0})}_{\rightarrow}, \\ 
 \frac{d}{ds}e^{i\int^{t}_{s}d\tau L(\tau,t_{0})}_{\leftarrow} &=& e^{i\int^{t}_{s}d\tau L(\tau,t_{0})}_{\leftarrow}(-iL(s,t_{0})), \\
 \frac{d}{dt}e^{i\int^{t}_{s}d\tau L(\tau,t_{0})}_{\rightarrow}e^{-i\int^{t}_{s}d\tau L(\tau,t_{0})}_{\leftarrow}
 &=& \frac{d}{ds}e^{-i\int^{t}_{s}d\tau L(\tau,t_{0})}_{\leftarrow}e^{i\int^{t}_{s}d\tau L(\tau,t_{0})}_{\rightarrow}
 =0. \label{eqn:T.O.bibun}
\end{eqnarray}
From Eq. (\ref{eqn:T.O.bibun}), we can derive the identities
\begin{eqnarray}
   e^{i\int^{t}_{s}d\tau L(\tau,t_{0})}_{\rightarrow}
   e^{-i\int^{t}_{s}d\tau L(\tau,t_{0})}_{\leftarrow}
 = e^{-i\int^{t}_{s}d\tau L(\tau,t_{0})}_{\leftarrow}
   e^{i\int^{t}_{s}d\tau L(\tau,t_{0})}_{\rightarrow}
 = 1. \label{eqn:normali}
\end{eqnarray}

The following equations are derived similarly to those derived in the previous section.
Corresponding to Eqs. (\ref{eqn:P}) and (\ref{eqn:Q}), 
we obtain 
\begin{eqnarray}
  \frac{d}{dt}e^{i\int^{t}_{t_{0}}ds L(s,t_{0})}_{\rightarrow}P 
&=& e^{i\int^{t}_{t_{0}}ds L(s,t_{0})}_{\rightarrow}PiL(t,t_{0})P 
    +e^{i\int^{t}_{t_{0}}ds L(s,t_{0})}_{\rightarrow}QiL(t,t_{0})P, \nonumber \\
\label{eqn:P2}\\
  \frac{d}{dt}e^{i\int^{t}_{t_{0}}ds L(s,t_{0})}_{\rightarrow}Q 
&=& e^{i\int^{t}_{t_{0}}ds L(s,t_{0})}_{\rightarrow}PiL(t,t_{0})Q 
    +e^{i\int^{t}_{t_{0}}ds L(s,t_{0})}_{\rightarrow}QiL(t,t_{0})Q. \nonumber \\
\label{eqn:Q2}
\end{eqnarray}
Equation (\ref{eqn:Q2}) gives us 
\begin{eqnarray}
  e^{i\int^{t}_{t_{0}}ds L(s,t_{0})}_{\rightarrow}Q 
&=& Qe^{i\int^{t}_{t_{0}}ds L(s,t_{0})Q}_{\rightarrow} 
    + \int^{t}_{t_{0}}ds e^{i\int^{s}_{t_{0}}d\tau 
    L(\tau,t_{0})}_{\rightarrow}PiL(s,t_{0})Q
    e^{i\int^{t}_{s}d\tau L(\tau,t_{0})Q}_{\rightarrow} \nonumber \\
&=& Qe^{i\int^{t}_{t_{0}}ds L(s,t_{0})Q}_{\rightarrow} \nonumber \\
&& + e^{i\int^{t}_{t_{0}}ds L(s,t_{0})}_{\rightarrow}(P+Q)
    \int^{t}_{ t_{0} }ds e^{-i\int^{t}_{t_{0}}d\tau L(\tau,t_{0})}_{\leftarrow}
    e^{i\int^{s}_{t_{0}}d\tau L(\tau,t_{0})}_{\rightarrow} \nonumber \\
&&\times  PiL(s,t_{0})Qe^{i\int^{t}_{s}d\tau L(\tau,t_{0})Q}_{\rightarrow} \nonumber \\
&=& \{ Qe^{i\int^{t}_{t_{0}}ds L(s,t_{0})Q}_{\rightarrow} 
    + e^{i\int^{t}_{t_{0}}ds L(s,t_{0})}_{\rightarrow}P\Sigma_{\rm ex}(t,t_{0}) \}
    \frac{1}{1-\Sigma_{\rm ex}(t,t_{0})}, \nonumber \\
\label{eqn:tcl}
\end{eqnarray}
where
\begin{eqnarray}
  \Sigma_{\rm ex}(t,t_{0}) = \int^{t}_{ t_{0} }ds e^{-i\int^{t}_{t_{0}}d\tau L(\tau,t_{0})}_{\leftarrow}
e^{i\int^{s}_{t_{0}}d\tau L(\tau,t_{0})}_{\rightarrow}
PiL(s,t_{0})Qe^{i\int^{t}_{s}d\tau L(\tau,t_{0})Q}_{\rightarrow}. \nonumber \\
\end{eqnarray}
Therefore, we can obtain the following equation:
\begin{eqnarray}
  \frac{d}{dt}O(t) 
&=& e^{i\int^{t}_{t_{0}}ds L(s,t_{0})}_{\rightarrow}PiL(t,t_{0})O(t_{0}) \nonumber \\
&&          + e^{i\int^{t}_{t_{0}}ds L(s,t_{0})}_{\rightarrow}
          P\Sigma_{\rm ex}(t,t_{0})\frac{1}{1-\Sigma_{\rm ex}(t,t_{0})}iL(t,t_{0})O(t_{0}) \nonumber\\
&&+ Qe^{i\int^{t}_{t_{0}}ds L(s,t_{0})Q}_{\rightarrow}\frac{1}{1-\Sigma_{\rm ex}(t,t_{0})}iL(t,t_{0})O(t_{0}).
\end{eqnarray}
This equation is equivalent to the Heisenberg equation, and 
corresponds to Eq. (\ref{eqn:genmitu-nt}) in \S 2.
When we ignore the explicit time dependence of the Hamiltonian, 
this equation agrees with Eq. (\ref{eqn:genmitu-nt}).

Now we define the projection operator.
As in the previous section, we consider the case in which 
the total Hamiltonian can be divided into three parts, 
the system (S), the environment (E), and the interaction (I) between 
the system and the environment:
\begin{eqnarray}
  H(t,t)  = H_{S}(t,t) + H_{E}(t,t)  + H_{I}(t,t) .
\end{eqnarray}
The self-interaction of the system and/or the environment 
can also be included in $H_{I}(t,t)$.
Furthermore, we assume the same form of the initial density matrix, 
given at the initial time $t_{0}$, as in Eq. (\ref{eqn:ini-density}).
We then define the projection operator as 
\begin{eqnarray}
  PO = {\rm Tr}_{E}[\rho_{E} O] \equiv \langle O \rangle_{E} 
\label{eqn:ex-pro2}
\end{eqnarray}
for any operator $O$.
From the nature of the projection operator, 
we obtain the relations
\begin{eqnarray}
  PL_{S}(t,t_{0}) &=& L_{S}(t,t_{0})P, ~QL_{S}(t,t_{0}) = L_{S}(t,t_{0})Q, \\
  L_{E}(t,t_{0})P &=& 0, ~~~~L_{E}(t,t_{0})Q = L_{E}(t,t_{0}), 
\end{eqnarray}
where $L_{a}(t,t_{0}) O = [H_{a}(t,t_{0}),O]$ for $a = S,E,I$.
Using these properties of the projection operator, 
we have 
\begin{eqnarray}
  QL_{0}(t,t_{0})Q &=& QL_{0}(t,t_{0}), \label{eqn:nrel2} \\
%  Qe^{iL_{0}t}Q &=& Qe^{iQL_{0}Qt}Q = Qe^{iL_{0}t},\label{eqn:nreld2}
Qe^{i\int^{t}_{t_{0}}ds L_{0}(s,t_{0})}_{\rightarrow(\leftarrow)}Q 
&=& Qe^{i\int^{t}_{t_{0}}ds QL_{0}(s,t_{0})Q}_{\rightarrow(\leftarrow)}Q 
= Qe^{i\int^{t}_{t_{0}}ds L_{0}(s,t_{0})}_{\rightarrow(\leftarrow)},\label{eqn:nreld2}
\end{eqnarray}
where $L_{0}(t,t_{0}) = L_{S}(t,t_{0}) + L_{E}(t,t_{0})$.

Next, 
$\Sigma_{\rm ex}(t,t_{0})$ is expressed as 
\begin{eqnarray}
  \Sigma_{\rm ex}(t,t_{0}) 
&=& \int^{t}_{ t_{0} }ds e^{-i\int^{t}_{t_{0}}d\tau L(\tau,t_{0})}_{\leftarrow}
    e^{i\int^{s}_{t_{0}}d\tau L(\tau,t_{0})}_{\rightarrow}
    PiL(s,t_{0})Qe^{i\int^{t}_{s}d\tau L(\tau,t_{0})Q}_{\rightarrow} \nonumber \\
&=& Q
    - e^{-i\int^{t}_{t_{0}}ds L(s,t_{0})}_{\leftarrow}Q
    e^{i\int^{t}_{t_{0}}ds L(s,t_{0})Q}_{\rightarrow} \nonumber \\
&=& Q 
    - U_{0}^{-1}(t,t_{0}){\cal C}_{\rm ex}(t,t_{0})Q
    {\cal D}_{\rm ex}(t,t_{0})U_{0}(t,t_{0}) \nonumber \\
&=& Q - U_{0}^{-1}(t,t_{0})QU_{0}(t,t_{0}) 
    - U_{0}^{-1}(t,t_{0})({\cal C}_{\rm ex}(t,t_{0})-1)Q
    U_{0}(t,t_{0}) \nonumber \\    
&&  - U_{0}^{-1}(t,t_{0})Q({\cal D}_{\rm ex}(t,t_{0})-1)
    U_{0}(t,t_{0}) \nonumber \\
&&  - U_{0}^{-1}(t,t_{0})({\cal C}_{\rm ex}(t,t_{0})-1)Q
    ({\cal D}_{\rm ex}(t,t_{0})-1)U_{0}(t,t_{0}).
\end{eqnarray}
Here, we have introduced the following functions:
\begin{eqnarray}
 {\cal C}_{\rm ex}(t,t_{0}) 
&=& 1 
+ \sum^{\infty}_{n=1}(-i)^n\int^{t}_{t_{0}}dt_{1}
\int^{t_{1}}_{t_{0}}dt_{2}\cdots\int^{t_{n-1}}_{t_{0}}dt_{n} \nonumber \\
&&\times \breve{L}^{\rm ex}_{I}(t_{1},t_{0})
\breve{L}^{\rm ex}_{I}(t_{2},t_{0}) \cdots 
\breve{L}^{\rm ex}_{I}(t_{n},t_{0}), \nonumber \\
\\
 {\cal D}_{\rm ex}(t,t_{0}) &=& 1 + \sum^{\infty}_{n=1}i^n\int^{t}_{t_{0}}dt_{1}\int^{t_{1}}_{t_{0}}dt_{2}\cdots\int^{t_{n-1}}_{t_{0}}dt_{n} Q\breve{L}^{Q~{\rm ex}}_{I}(t_{n},t_{0}) \nonumber \\
&&\times Q\breve{L}^{Q~{\rm ex}}_{I}(t_{n-1},t_{0}) \cdots Q\breve{L}^{Q~{\rm ex}}_{I}(t_{1},t_{0}), \nonumber \\
\end{eqnarray}
where
\begin{eqnarray}
  U_{0}(t,t_{0}) 
&=&  e^{i\int^{t}_{t_{0}}ds L_{0}(s,t_{0})}_{\rightarrow}, \\
  \breve{L}^{\rm ex}_{I}
(t,t_{0}) &=& U_{0}(t,t_{0})L_{I}(t,t_{0})U_{0}^{-1}(t,t_{0}), \\
  \breve{L}^{Q~{\rm ex}}_{I}(t,t_{0}) &=& U_{0}(t,t_{0})L_{I}(t,t_{0})QU_{0}^{-1}(t,t_{0}). 
\end{eqnarray}
%Here, $L_{0}(s,t_{0})$ has explicit time dependence 
%which is different from U-S projection operator method.
By using the above functions and mathematical induction, 
we obtain 
\begin{eqnarray}
\lefteqn{   \frac{d}{dt}O(t) } && \nonumber \\
&=&  e^{i\int^{t}_{t_{0}}ds L(s,t_{0})}_{\rightarrow}
     PiL(t,t_{0})O(t_{0}) \nonumber \\
&&  -e^{i\int^{t}_{t_{0}}ds L(s,t_{0})}_{\rightarrow}
     PU_{0}^{-1}(t,t_{0}){\cal C}_{\rm ex}(t,t_{0})Q    
    \frac{1}{1+({\cal C}_{\rm ex}(t,t_{0})-1)Q}
     U_{0}(t,t_{0})iL(t,t_{0})O(t_{0}) \nonumber \\
&&  + Qe^{i\int^{t}_{t_{0}}ds L(s,t_{0})Q}_{\rightarrow}
    \frac{1}{1-\Sigma_{\rm ex}(t,t_{0})}iL(t,t_{0})O(t_{0}).
    \label{eqn:t-genmitu-2}
\end{eqnarray}
This is the final expression of the Heisenberg equation of motion 
with time-dependent Hamiltonian.
Perturbative expansion up to first order in 
$P\Sigma_{\rm ex}(t,t_{0})/(1-\Sigma_{\rm ex}(t,t_{0}))$
leads to 
\begin{eqnarray}
\lefteqn{   \frac{d}{dt}O(t) }&&\nonumber \\
&=& e^{i\int^{t}_{t_{0}}ds L(s,t_{0}) }_{\longrightarrow}
            PU_{0}^{-1}(t,t_{0})PU_{0}(t,t_{0})
            iL(t,t_{0})O(t_{0}) \nonumber \\
&& +e^{i\int^{t}_{t_{0}}ds L(s,t_{0}) }_{\longrightarrow}
               PU_{0}^{-1}(t,t_{0})P\int^{t}_{t_{0}}ds 
            U_{0}(s,t_{0})iL_{I}(s,t_{0})U^{-1}_{0}(s,t_{0})Q 
            U_{0}(t,t_{0}) iL(t,t_{0})O(t_{0}) \nonumber \\
&& +  Qe^{i\int^{t}_{t_{0}}ds L(s,t_{0})Q}_{\rightarrow}
            \frac{1}{1-\Sigma_{\rm ex}(t,t_{0})}iL(t,t_{0})O(t_{0}). \label{eqn:t-res}
\end{eqnarray}
Here, we do not expand the third term on the r.h.s.~of Eq. (\ref{eqn:t-res}).
This equation is the TCL equation in the case of a time-dependent Hamiltonian, 
and 
it becomes the same equation as Eq. (\ref{eqn:t-i-res}) 
when we ignore the explicit time dependence in the Hamiltonian.
The TC equation can be obtained as in Appendix C.

\section{The Damped harmonic oscillator model}

In this section, we apply our improved projection operator method to 
the exactly solvable model called 
the damped harmonic oscillator model,\cite{ref:8} 
which has no explicit time dependence.
The Hamiltonian is 
\begin{eqnarray}
H = \sum_{k}\left\{
\Omega_{k}(A^{\dagger}_{k}A_{k}-B^{\dagger}_{k}B_{k}) 
+ i\Gamma_{k}(A^{\dagger}_{k}B^{\dagger}_{k}-A_{k}B_{k})
\right\}.
\end{eqnarray}
The $k$ dependence of $\Gamma_{k}$ is such that 
$\displaystyle \sum_{k}\Gamma_{k}$ is  finite.
We impose the following commutation relation:
\begin{eqnarray}
[A_{i},A^{\dagger}_{j}] = [B_{i},B^{\dagger}_{j}] = \delta_{i,j},~~~~~~
[A_{i},B_{j}] = 0.
\end{eqnarray}
The time development of the operator $A^{\dagger}_{k}(t)$ 
is exactly solved as 
\begin{eqnarray}
A^{\dagger}_{k}(t)
 = e^{i\Omega_{k} t}
\left(
{\rm cosh}(\Gamma_{k}t)A^{\dagger}_{k}(0) 
+ {\rm sinh}(\Gamma_{k}t)B_{k}(0)
\right).
\end{eqnarray}
This is a solution of the following equation of motion:
\begin{eqnarray}
\frac{d}{dt}\langle A^{\dagger}_{k}(t) \rangle
= i\Omega_{k}\langle A^{\dagger}_{k}(t) \rangle 
  + \Gamma_{k}{\rm tanh}(\Gamma_{k} t)\langle A^{\dagger}_{k}(t) \rangle
  + e^{i\Omega_{k} t}\Gamma_{k} \frac{1}{{\rm cosh}(\Gamma_{k} t)}\langle B_{k}(0) \rangle.\nonumber \\
 \label{eqn:DHO-gen}
\end{eqnarray}
Here, $\langle~\rangle$ represents the expectation value taken with respect to 
an arbitrary density matrix $\rho$ that 
satisfies the condition (\ref{eqn:ini-density}).

Now, we apply our method to this model, and compare the result with 
the exact equation (\ref{eqn:DHO-gen}).
We consider the degree of freedom of the system to be $A_{k}$ 
and that of the environment to be $B_{k}$.
The Hamiltonian is divided into three parts:
\begin{eqnarray}
H_{S} &=&  \sum_{k}\Omega_{k} A^{\dagger}_{k}A_{k},  \\ 
H_{E} &=&  -\sum_{k}\Omega_{k} B^{\dagger}_{k}B_{k}, \\
H_{I} &=&  \sum_{k}i\Gamma_{k}(A^{\dagger}_{k}B^{\dagger}_{k}-A_{k}B_{k}).
\end{eqnarray}
The projection operator is defined in Eq. (\ref{eqn:ex-pro}).
When we substitute $t_{0}=0$ and $O(0)=A^{\dagger}_{k}$ 
into Eq. (\ref{eqn:t-i-res}), we have
\begin{eqnarray}
\frac{d}{dt}\langle A^{\dagger}_{k}(t) \rangle
= i\Omega_{k} \langle A^{\dagger}_{k}(t) \rangle + \Gamma_{k}^2 t \langle A^{\dagger}_{k}(t) \rangle
  + e^{i\Omega_{k} t}\Gamma_{k} \langle B_{k}(0) \rangle.
\end{eqnarray}
This equation is identical with Eq. (\ref{eqn:DHO-gen}) 
when we expand ${\rm tanh}(\Gamma_{k} t)$ and $1/{\rm cosh(\Gamma_{k} t)}$ 
to lowest order in $\Gamma_{k}$.
Furthermore, higher-order contributions are easily calculable.
By using Eq. (\ref{eqn:kouji}) in Appendix D instead of Eq. (\ref{eqn:t-i-res}), 
we can obtain the equation of motion with a higher-order term:
\begin{eqnarray}
\frac{d}{dt}\langle A^{\dagger}_{k}(t) \rangle
= i\Omega_{k} \langle A^{\dagger}_{k}(t) \rangle + \Gamma_{k}^2 t \langle A^{\dagger}_{k}(t) \rangle
  + e^{i\Omega_{k} t}\Gamma_{k} \left( 1-\frac{1}{2}\Gamma_{k}^2 t^2 \right) \langle B_{k}(0) \rangle.
\end{eqnarray}
This equation also coincides with Eq. (\ref{eqn:DHO-gen}) with the appropriate 
perturbative expansion of $\Gamma_{k}$.

Now, we apply this model to the previous projection operator method.
The approximate equation in that method \cite{ref:1} which 
corresponds to Eq. (\ref{eqn:t-i-res}) is 
\begin{eqnarray}
  \frac{d}{dt}O(t) 
&=& e^{iL(t-t_{0})}PiLO(t_{0}) \nonumber \\
   && + e^{iL(t-t_{0})}\int^{t}_{t_{0}}ds e^{-iL_{S}(s-t_{0})}PiL_{I}e^{iL_{0}(s-t_{0})}QiLO(t_{0}) \nonumber \\
   && + Qe^{iLQ(t-t_{0})}iLO(t_{0}). 
\label{eqn:s-h}
\end{eqnarray}
Here, we use the same initial density matrix and projection operator.
Now, the equation of motion for $A^{\dagger}_{k}$ is 
\begin{eqnarray}
\frac{d}{dt}\langle A^{\dagger}_{k}(t) \rangle
= i\Omega_{k} \langle A^{\dagger}_{k}(t) \rangle + \Gamma_{k}^2 t \langle A^{\dagger}_{k}(t) \rangle
  + \Gamma_{k} \langle B_{k}(0) \rangle.
\end{eqnarray}
It seems that the factor $e^{i\Omega_{k} t}$ in the third term 
on the r.h.s.~of Eq.~(\ref{eqn:DHO-gen}) cannot be reproduced.
However, Eq. (\ref{eqn:s-h}) gives consistent results 
only when we use a restricted density matrix that satisfies 
the condition $PL_{E} = 0$.
This condition implies 
\begin{eqnarray}
PL_{E}B_{k}(0) = 0 \longrightarrow \langle B_{k}(0) \rangle=0.
\end{eqnarray}
Therefore, the above equation is consistent with the exact result.
From this argument, 
we can see that the previous projection operator method is 
valid only for a restricted initial state.

\section{Conclusions and discussion}

We have given a new expansion of the Heisenberg equation of motion with 
a projection operator in the two cases of 
Hamiltonian both with and without explicit time dependence.
In our method, one can prepare more general initial states 
which are forbidden in the previously applied projection operator method.
Until now, the projection operator has been chosen as Eq. (\ref{eqn:ex-pro}).
However, if the condition $QL_{0}Q=QL_{0}$ 
(in other words, $PL_{0}P=L_{0}P$) is satisfied, 
one can use Eqs. (\ref{eqn:genmitu-nt-2}) and (\ref{eqn:t-genmitu-2}) 
for any kind of projection operators.

Recently, 
a similar expansion method with a projection operator 
was proposed by Uchiyama and Shibata.\cite{ref:9}
In their method, one can expand the Heisenberg equation 
without such a restriction on projection operators as $QL_{0}Q=QL_{0}$, 
which is needed in our method.
However, they impose some conditions on the operator $O(t_{0})$ 
whose time evolution we want to calculate.
The first condition is that 
the time evolution with the unperturbed Hamiltonian should be 
solved as 
\begin{eqnarray}
e^{iL_{0}(t-t_{0})}O(t_{0}) = f(t,t_{0})O(t_{0}), 
\end{eqnarray}
where $f(t,t_{0})$ is a c-number.
The second condition is  
\begin{eqnarray}
PO(t_{0}) = O(t_{0}).
\end{eqnarray}
This condition implies that 
we cannot calculate the time evolution of environment operators, 
when we use the definition (\ref{eqn:ex-pro}).
However, it is possible to formulate the method without the above conditions.
This extended Uchiyama-Shibata projection operator method 
is explained in Appendix E.

\appendix

\section{The Definition of the Operators ${\cal C}$ and ${\cal D}$}

We define the operators ${\cal C}(t,t_{0})$ and ${\cal D}(t,t_{0})$ as
\begin{eqnarray}
   e^{-iL(t-t_{0})} &=& e^{-iL_{0}(t-t_{0})}{\cal C}(t,t_{0}), \nonumber\\
  {\cal C}(t,t_{0}) &=& e^{iL_{0}(t-t_{0})}e^{-iL(t-t_{0})},\\
   e^{iQLQ(t-t_{0})} &=& {\cal D}(t,t_{0})e^{iQL_{0}Q(t-t_{0})}, \nonumber\\
   {\cal D}(t,t_{0}) &=& e^{iQLQ(t-t_{0})}e^{-iQL_{0}Q(t-t_{0})}. 
\end{eqnarray}
These operators satisfy the following differential equations: 
\begin{eqnarray}
    \frac{d}{dt}{\cal C}(t,t_{0}) 
&=& e^{iL_{0}(t-t_{0})}(iL_{0}-iL)e^{-iL(t-t_{0})} \nonumber \\
&=& -i\breve{L}_{I}(t,t_{0}){\cal C}(t,t_{0}),  \\
    \frac{d}{dt}{\cal D}(t,t_{0}) 
&=& e^{iQLQ(t-t_{0})}Q(iL-iL_{0})Qe^{-iQL_{0}Q(t-t_{0})} \nonumber \\
&=& {\cal D}(t,t_{0})Qi\breve{L}^{Q}_{I}(t,t_{0}),
\end{eqnarray}
where
\begin{eqnarray}
  \breve{L}_{I}(t,t_{0}) &=& e^{iL_{0}(t-t_{0})}L_{I}e^{-iL_{0}(t-t_{0})}, \\
  \breve{L}^{Q}_{I}(t,t_{0}) &=& e^{iL_{0}(t-t_{0})}L_{I}Qe^{-iL_{0}(t-t_{0})}.
\end{eqnarray}
From these differential equations, we obtain
\begin{eqnarray}
  {\cal C}(t,t_{0}) &=& 1 + \sum^{\infty}_{n=1}(-i)^n\int^{t}_{t_{0}}dt_{1}\int^{t_{1}}_{t_{0}}dt_{2}\cdots\int^{t_{n-1}}_{t_{0}}dt_{n} \breve{L}_{I}(t_{1}-t_{0})\breve{L}_{I}(t_{2}-t_{0}) \nonumber \\
  && \times \cdots \breve{L}_{I}(t_{n}-t_{0}), \\
  {\cal D}(t,t_{0}) &=& 1 + \sum^{\infty}_{n=1}i^n\int^{t}_{t_{0}}dt_{1}\int^{t_{1}}_{t_{0}}dt_{2}\cdots\int^{t_{n-1}}_{t_{0}}dt_{n} Q\breve{L}^{Q}_{I}(t_{n}-t_{0})Q
  \breve{L}^{Q}_{I}(t_{n-1}-t_{0}) \nonumber \\
  &&\times \cdots Q\breve{L}^{Q}_{I}(t_{1}-t_{0}).
\end{eqnarray}

Similarly, ${\cal C}_{\rm ex}(t,t_{0})$ and ${\cal D}_{\rm ex}(t,t_{0})$ 
are defined as 
\begin{eqnarray}
   e^{-i\int^{t}_{t_{0}}ds L(s,t_{0})}_{\leftarrow} &=& U_{0}^{-1}(t,t_{0}){\cal C}_{\rm ex}(t,t_{0}), \nonumber\\
  {\cal C}_{\rm ex}(t,t_{0}) &=& U_{0}(t,t_{0})e^{-i\int^{t}_{t_{0}}ds L(s,t_{0})}_{\leftarrow}, \\
   e^{i\int^{t}_{t_{0}}ds QL(s,t_{0})Q}_{\rightarrow} &=& {\cal D}_{\rm ex}(t,t_{0})e_{\rightarrow}^{i\int^{t}_{t_{0}}ds QL_{0}(s,t_{0})Q }, \nonumber\\
   {\cal D}_{\rm ex}(t,t_{0}) &=& e^{i\int^{t}_{t_{0}}ds QL(s,t_{0})Q}_{\rightarrow}e_{\leftarrow}^{-i\int^{t}_{t_{0}}ds QL_{0}(s,t_{0})Q }. 
\end{eqnarray}
Differential equations can be constructed again, and are solved to yield 
\begin{eqnarray}
    {\cal C}_{\rm ex}(t,t_{0}) &=& 1 + \sum^{\infty}_{n=1}(-i)^n\int^{t}_{t_{0}}dt_{1}\int^{t_{1}}_{t_{0}}dt_{2}\cdots\int^{t_{n-1}}_{t_{0}}dt_{n} \breve{L}^{\rm ex}_{I}(t_{1},t_{0})\breve{L}^{\rm ex}_{I}(t_{2},t_{0}) \nonumber \\
  && \times \cdots \breve{L}^{\rm ex}_{I}(t_{n},t_{0}), \\
  {\cal D}_{\rm ex}(t,t_{0}) &=& 1 + \sum^{\infty}_{n=1}i^n\int^{t}_{t_{0}}dt_{1}\int^{t_{1}}_{t_{0}}dt_{2}\cdots\int^{t_{n-1}}_{t_{0}}dt_{n} Q\breve{L}^{Q~{\rm ex}}_{I}(t_{n},t_{0})Q
  \breve{L}^{Q~{\rm ex}}_{I}(t_{n-1},t_{0}) \nonumber \\
  &&\times \cdots Q\breve{L}^{Q~{\rm ex}}_{I}(t_{1},t_{0}),
\end{eqnarray}
where
\begin{eqnarray}
  \breve{L}^{\rm ex}_{I}(t,t_{0}) &=& U_{0}(t,t_{0})L_{I}(t,t_{0})U_{0}^{-1}(t,t_{0}), \\
  \breve{L}^{Q~{\rm ex}}_{I}(t,t_{0}) &=& U_{0}(t,t_{0})L_{I}(t,t_{0})QU_{0}^{-1}(t,t_{0}). 
\end{eqnarray}

\section{The Transformation of the Operator $\Sigma (t,t_{0})$}

The operator $\Sigma (t,t_{0})$ in \S 2 can be expressed as
\begin{eqnarray}
  \Sigma(t,t_{0}) &=& \int^{t}_{ t_{0} }ds e^{-iL(t-s)}
PiLQe^{iL(t-s)Q} \nonumber \\
            &=& -\int^{t}_{ t_{0} }ds e^{-iL(t-s)}
P\frac{d}{ds}e^{iL(t-s)Q} \nonumber \\
            &=& - \Big\{ P - e^{-iL(t-t_{0})}Pe^{iL(t-t_{0})Q}
-\int^{t}_{t_{0}}ds e^{-iL(t-s)}iLPe^{iL(t-s)Q} \Big\} \nonumber \\
            &=& - \Big\{ P - e^{-iL(t-t_{0})}Pe^{iL(t-t_{0})Q}
-\int^{t}_{t_{0}}ds \frac{d}{ds}(e^{-iL(t-s)}e^{iL(t-s)Q}) \Big\} \nonumber \\
            &=& - \{ P - e^{-iL(t-t_{0})}Pe^{iL(t-t_{0})Q}
- 1 + e^{-iL(t-t_{0})}e^{iL(t-t_{0})Q} \} \nonumber \\
            &=& Q - e^{-iL(t-t_{0})}Qe^{iL(t-t_{0})Q} \nonumber \\
            &=& Q - e^{-iL_{0}(t-t_{0})}{\cal C}(t,t_{0})
                   Q{\cal D}(t,t_{0})e^{iQL_{0}Q(t-t_{0})}.
\end{eqnarray}
Similarly, $\Sigma_{\rm ex} (t,t_{0})$ in \S 3 is given by 
\begin{eqnarray}
  \Sigma_{\rm ex}(t,t_{0}) 
&=& \int^{t}_{ t_{0} }ds e^{-i\int^{t}_{t_{0}}d\tau L(\tau,t_{0})}_{\leftarrow}
e^{i\int^{s}_{t_{0}}d\tau L(\tau,t_{0})}_{\rightarrow}
PiL(s,t_{0})Qe^{i\int^{t}_{s}d\tau L(\tau,t_{0})Q}_{\rightarrow} \nonumber \\
      &=& -\int^{t}_{ t_{0} }ds e^{-i\int^{t}_{t_{0}}d\tau L(\tau,t_{0})}_{\leftarrow}
e^{i\int^{s}_{t_{0}}d\tau L(\tau,t_{0})}_{\rightarrow}
P\frac{d}{ds}e^{i\int^{t}_{s}d\tau L(\tau,t_{0})Q}_{\rightarrow} \nonumber \\
            &=& -e^{-i\int^{t}_{t_{0}}ds L(s,t_{0})}_{\leftarrow} \{
e^{i\int^{t}_{t_{0}}ds L(s,t_{0})}_{\rightarrow}P
-Pe^{i\int^{t}_{t_{0}}ds L(s,t_{0})Q}_{\rightarrow} \nonumber \\
&& -\int^{t}_{t_{0}}ds e^{i\int^{s}_{t_{0}}d\tau L(\tau,t_{0})}_{\rightarrow}
iL(s,t_{0})Pe^{i\int^{t}_{s}d\tau L(\tau,t_{0})Q}_{\rightarrow} \} \nonumber \\
            &=& -e^{-i\int^{t}_{t_{0}}ds L(s,t_{0})}_{\leftarrow} \Big\{
e^{i\int^{t}_{t_{0}}ds L(s,t_{0})}_{\rightarrow}P
-Pe^{i\int^{t}_{t_{0}}ds L(s,t_{0})Q}_{\rightarrow} \nonumber \\
&&-\int^{t}_{t_{0}}ds \frac{d}{ds}\Big( e^{i\int^{s}_{t_{0}}ds L(s,t_{0})}_{\rightarrow}
e^{i\int^{t}_{s}ds L(s,t_{0})Q}_{\rightarrow} \Big) \Big\} \nonumber \\
            &=& -e^{-i\int^{t}_{t_{0}}ds L(s,t_{0})}_{\leftarrow} \Big\{
e^{i\int^{t}_{t_{0}}ds L(s,t_{0})}_{\rightarrow}P
-Pe^{i\int^{t}_{t_{0}}ds L(s,t_{0})Q}_{\rightarrow} \nonumber \\
&&-e^{i\int^{t}_{t_{0}}ds L(s,t_{0})}_{\rightarrow}
+e^{i\int^{t}_{t_{0}}ds L(s,t_{0})Q}_{\rightarrow} \Big\} \nonumber \\
            &=& e^{-i\int^{t}_{t_{0}}ds L(s,t_{0})}_{\leftarrow}e^{i\int^{t}_{t_{0}}ds L(s,t_{0})}_{\rightarrow}Q
- e^{-i\int^{t}_{t_{0}}ds L(s,t_{0})}_{\leftarrow}Qe^{i\int^{t}_{t_{0}}ds L(s,t_{0})Q}_{\rightarrow} \nonumber \\
            &=& Q
- e^{-i\int^{t}_{t_{0}}ds L(s,t_{0})}_{\leftarrow}
  Qe^{i\int^{t}_{t_{0}}ds L(s,t_{0})Q}_{\rightarrow} \nonumber \\
            &=& Q
- U_{0}^{-1}(t,t_{0}){\cal C}_{\rm ex}(t,t_{0})
  Q{\cal D}_{\rm ex}(t,t_{0})e_{\rightarrow}^{i\int^{t}_{t_{0}}ds QL_{0}(s,t_{0})Q }.
\end{eqnarray}
To allow for simultaneous discussion, we introduce the following notation:
\begin{eqnarray}
{\cal C}(t) 
&=&  { e^{-iL_{0}(t-t_{0})}{\cal C}(t,t_{0})e^{iL_{0}(t-t_{0})}~~~~~{\rm for~the~case~of~\S~2}
\atopwithdelims\{. 
U_{0}^{-1}(t,t_{0}){\cal C}_{\rm ex}(t,t_{0})U_{0}(t,t_{0})~~~~{\rm for~the~case~of~\S~3},  } \\
{\cal D}(t) 
&=&  { e^{-iL_{0}(t-t_{0})}{\cal D}(t,t_{0})e^{iL_{0}(t-t_{0})}~~~~~{\rm for~the~case~of~\S~2}
\atopwithdelims\{. 
U_{0}^{-1}(t,t_{0}){\cal D}_{\rm ex}(t,t_{0})U_{0}(t,t_{0})~~~{\rm for~the~case~of~\S~3},  } \\
Q(t) 
&=&  { e^{-iL_{0}(t-t_{0})}Qe^{iL_{0}(t-t_{0})}~~~~~~~~~~~~{\rm for~the~case~of~\S~2}
\atopwithdelims\{. 
U_{0}^{-1}(t,t_{0})QU_{0}(t,t_{0})~~~~~~~~~~~~{\rm for~the~case~of~\S~3}.  }
\end{eqnarray}
Using the mathematical induction, we confirm 
the following relation:
\begin{eqnarray}
\lefteqn{     P\Sigma(t,t_{0})(Q\Sigma(t,t_{0}))^n  }&&\nonumber \\
&=&  [(-1)^{n-1}P\{ Q(t)({\cal C}(t)-1) \}^n Q(t)
     +(-1)^{n-1}P\{ ({\cal C}(t) -1) Q(t)\}^{n+1}] \nonumber \\
&& + P\sum_{l=0}^{n-1}(-1)^l \{ Q(t)({\cal C}(t)-1) \}^l Q(t)({\cal D}(t)-1)
     (Q\Sigma(t,t_{0}))^{n-1-l} \nonumber \\
&& + P\sum_{l=0}^{n-1}(-1)^l \{ ({\cal C}(t)-1)Q(t) \}^{l+1} ({\cal D}(t)-1)
     (Q\Sigma(t,t_{0}))^{n-1-l} \nonumber \\ 
&& - P\sum_{l=0}^{n}(-1)^l \{ Q(t)({\cal C}(t)-1) \}^l Q(t)({\cal D}(t)-1)
     (Q\Sigma(t,t_{0}))^{n-l} \nonumber \\
&& - P\sum_{l=0}^{n}(-1)^l \{ ({\cal C}(t)-1)Q(t) \}^{l+1} ({\cal D}(t)-1)
     (Q\Sigma(t,t_{0}))^{n-l},
\end{eqnarray}
where $n$ is integer and $n\geq 1$.
The second and third terms in $P\Sigma(t,t_{0})(Q\Sigma(t,t_{0}))^{n}$ 
and the fourth and fifth terms in $P\Sigma(t,t_{0})(Q\Sigma(t,t_{0}))^{n-1}$
cancel.
The fourth and fifth terms in $P\Sigma(t,t_{0})(Q\Sigma(t,t_{0}))^{n}$ 
and the second and third terms in $P\Sigma(t,t_{0})(Q\Sigma(t,t_{0}))^{n+1}$
also cancel.
Therefore, only the first term survives.
As a result, all the terms including ${\cal D}(t)$ disappear.
Finally, noting the relation $\Sigma(t,t_{0})=\Sigma(t,t_{0})Q$, 
we find that $P\Sigma(t,t_{0})\frac{1}{1-\Sigma(t,t_{0})}$ can be expressed as follows:
\begin{eqnarray}
         P\Sigma(t,t_{0})\frac{1}{1-\Sigma(t,t_{0})} 
&=&  P\Sigma(t,t_{0})\frac{1}{1-Q\Sigma(t,t_{0})} \nonumber \\
&=&  P\Sigma(t,t_{0})\sum_{n=0}^{\infty}(Q\Sigma(t,t_{0}))^n \nonumber \\
&=&  -P\sum_{n=0}^{\infty}[ \{ -Q(t)({\cal C}(t)-1) \}^n Q(t)
     -\{ -({\cal C}(t)-1)Q(t) \}^{n+1}  ] \nonumber \\
&=&  -PQ(t)\frac{1}{1+({\cal C}(t)-1)Q(t)} 
     -P\frac{({\cal C}(t)-1)Q(t)}{1+({\cal C}(t)-1)Q(t)} \nonumber \\
&=&  -P{\cal C}(t)Q(t)\frac{1}{1+({\cal C}(t)-1)Q(t)}. 
\end{eqnarray}

\section{The Derivation of the Time-Convolution Equation}

Here, we derive the equation with a time-convolution integral.
In this appendix, we consider the case of 
a Hamiltonian without explicit time dependence.
The discussion, however,  can also be applied to the case of 
a Hamiltonian with explicit time dependence.
We substitute Eq. (\ref{eqn:TC-dainyuu}), 
(instead of (\ref{eqn:TCL})) into Eq. (\ref{eqn:P+Q}), and we obtain 
\begin{eqnarray}
  \frac{d}{dt}O(t) 
&=& e^{iL(t-t_{0})}PiLO(t_{0})+\int^{t}_{t_{0}}ds e^{iL(s-t_{0})}PiLQe^{iLQ(t-s)}iLO(t_{0}) \nonumber \\
&& + Qe^{iLQ(t-t_{0})}iLO(t_{0}) \nonumber \\
&=& e^{iL(t-t_{0})}PiLO(t_{0})+\int^{t}_{t_{0}}ds e^{iL(t-s)}PiLQe^{iLQ(s-t_{0})}iLO(t_{0}) \nonumber \\
&& + Qe^{iLQ(t-t_{0})}iLO(t_{0}). \label{eqn:genmitu-tc1}
\end{eqnarray}
This equation is equivalent to the Heisenberg equation.
For the existence of the operator $e^{iL(t-s)}$,
the second term on the r.h.s.~of the equation has a time-convolution integral.
Therefore, this equation is a called a time-convolution (TC) equation.
This can be rewritten by using ${\cal D}(t,t_{0})$ as 
\begin{eqnarray}
  \frac{d}{dt}O(t) 
&=& e^{iL(t-t_{0})}PiLO(t_{0})+\int^{t}_{t_{0}}ds 
    e^{iL(t-s)}PiLQ{\cal D}(s,t_{0})e^{iQL_{0}Q(s-t_{0})}iLO(t_{0}) \nonumber \\
&&+ Qe^{iLQ(t-t_{0})}iLO(t_{0}). \label{eqn:genmitu-tc2}
\end{eqnarray}
By expanding ${\cal D}(t,t_{0})$ to lowest order, we have
\footnote{It may seem difficult to solve $e^{iQL_{0}Q(s-t_{0})}iLO(t_{0})$, 
because of the operator $Q$.
However, it can be solved using the following differential equation:
$\frac{d}{ds}e^{iQL_{0}Q(s-t_{0})}iLO(t_{0})=e^{iQL_{0}Q(s-t_{0})}iQL_{0}QiLO(t_{0})$.} 
\begin{eqnarray}
  \frac{d}{dt}O(t) 
&=& e^{iL(t-t_{0})}PiLO(t_{0})+\int^{t}_{t_{0}}ds 
    e^{iL(t-s)}PiLQe^{iQL_{0}Q(s-t_{0})}iLO(t_{0}) \nonumber \\
&&+ Qe^{iLQ(t-t_{0})}iLO(t_{0}). \label{eqn:ap-tc}
\end{eqnarray}
At this level of expansion, the difference between 
the equation in the improved projection operator method and 
that in the previous projection operator method cannot be observed, 
because this difference is included 
in the higher-order term in ${\cal D}(t,t_{0})$.

In the case of the TC equation, the expansion (\ref{eqn:genmitu-tc2}) is 
always valid, because no restriction is imposed on the projection operator.
This is different from the case of the TCL equation, 
for which the condition $QL_{0}Q=QL_{0}$ must be satisfied.

\section{The Derivation of Higher-Order Terms}

We expand 
${\cal C}(t,t_{0})/\{1+({\cal C}(t,t_{0})-1)\}$ 
to second order in the interaction:
\begin{eqnarray}
\lefteqn{{\cal C}(t,t_{0})
Q\frac{1}{1+({\cal C}(t,t_{0})-1)}Q} &&\nonumber \\
&=&
{\cal C}(t,t_{0})Q
\{ 
1
- ( {\cal C}(t,t_{0})-1 )Q 
+ ( {\cal C}(t,t_{0})-1 )Q( {\cal C}(t,t_{0})-1 )Q + \cdots 
\} \nonumber \\
&\sim& 
\Big\{
1 - i\int^{t}_{t_{0}}dt_{1}\breve{L}_{I}(t_{1}-t_{0})
+ (-i)^2\int^{t}_{t_{0}}dt_{1}\int^{t_{1}}_{t_{0}}dt_{2}
  \breve{L}_{I}(t_{1}-t_{0})\breve{L}_{I}(t_{2}-t_{0})
\Big\}Q \nonumber \\
&& \times
\{
1
+ i\int^{t}_{t_{0}}dt_{1}\breve{L}_{I}(t_{1}-t_{0})Q
- (-i)^2\int^{t}_{t_{0}}dt_{1}\int^{t_{1}}_{t_{0}}dt_{2}
  \breve{L}_{I}(t_{1}-t_{0})\breve{L}_{I}(t_{2}-t_{0})Q  \nonumber \\
&&+ (-i)^2\int^{t}_{t_{0}}dt_{1} \breve{L}_{I}(t_{1}-t_{0})Q
        \int^{t}_{t_{0}}dt_{2} \breve{L}_{I}(t_{2}-t_{0})Q
\} \nonumber \\
&\sim& 
Q-P\int^{t}_{t_{0}}dt_{1}i\breve{L}_{I}(t_{1}-t_{0})Q
+P\int^{t}_{t_{0}}dt_{1}\int^{t_{1}}_{t_{0}}dt_{2}
  i\breve{L}_{I}(t_{1}-t_{0})Pi\breve{L}_{I}(t_{2}-t_{0})Q \nonumber \\
&&-P\int^{t}_{t_{0}}dt_{1}\int^{t_{1}}_{t_{0}}dt_{2}
  i\breve{L}_{I}(t_{2}-t_{0})Qi\breve{L}_{I}(t_{1}-t_{0})Q. 
\end{eqnarray}
In substituting this into Eq. (\ref{eqn:genmitu-nt-2}), we have
\begin{eqnarray}
\lefteqn{  \frac{d}{dt}O(t) }&&\nonumber \\
   &=& e^{iL(t-t_{0})}Pe^{-iL_{0}(t-t_{0})}Pe^{iL_{0}(t-t_{0})}iLO(t_{0}) \nonumber \\
   && + e^{iL(t-t_{0})}Pe^{-iL_{0}(t-t_{0})}
      P\int^{t}_{t_{0}}ds 
      i\breve{L}_{I}(s-t_{0})Q
      e^{iL_{0}(t-t_{0})}iLO(t_{0}) \nonumber \\
   && - e^{iL(t-t_{0})}Pe^{-iL_{0}(t-t_{0})}
      P\int^{t}_{t_{0}}dt_{1}\int^{t_{1}}_{t_{0}}dt_{2}
      i\breve{L}_{I}(t_{1}-t_{0})
     Pi\breve{L}_{I}(t_{2}-t_{0})Q
      e^{iL_{0}(t-t_{0})}iLO(t_{0}) \nonumber \\
   && + e^{iL(t-t_{0})}Pe^{-iL_{0}(t-t_{0})}
      P\int^{t}_{t_{0}}dt_{1}\int^{t_{1}}_{t_{0}}dt_{2}
      i\breve{L}_{I}(t_{2}-t_{0})
     Qi\breve{L}_{I}(t_{1}-t_{0})Q
      e^{iL_{0}(t-t_{0})}iLO(t_{0}) \nonumber \\
   && + Qe^{iLQ(t-t_{0})}\frac{1}{1-\Sigma(t,t_{0})}iLO(t_{0}).
\label{eqn:kouji}
\end{eqnarray}
The third and fourth lines on the r.h.s.~of this equation are 
higher-order contributions.

\section{The Extended Uchiyama-Shibata Projection Operator Method}

Here we explain the extended 
Uchiyama-Shibata (U-S) projection operator method.
The time evolution of an operator can be written as 
\begin{eqnarray}
O(t) &=& e^{i\int^{t}_{t_{0}}dsL(s,t_{0})}_{\rightarrow}O(t_{0}) \nonumber \\
     &=& \hat{U}_{-}(t,t_{0})(U_{0}^{-1}(t,t_{0}))^{\dagger}O(t_{0}),
\end{eqnarray}
where
\begin{eqnarray}
\hat{U}_{-}(t,t_{0}) = e^{i\int^{t}_{t_{0}}ds \breve{L}_{I}(s,t_{0})}_{\rightarrow}.
\end{eqnarray}
Here $\breve{L}_{I}(t,t_{0})$ is the same as that defined in \S 2.
Now, we expand all the fields with creation and annihilation operators 
at $t_{0}$, and prepare the initial state at $t_{0}$.
The time evolution is determined by the following equation:
\begin{eqnarray}
\frac{d}{dt}O(t) 
&=& \frac{d}{dt}\hat{U}_{-}(t,t_{0})\cdot (U_{0}^{-1}(t,t_{0}))^{\dagger}O(t_{0}) 
    + \hat{U}_{-}(t,t_{0})\frac{d}{dt}(U_{0}^{-1}(t,t_{0}))^{\dagger}O(t_{0}). \nonumber \\
\label{eqn:t-bunri}
\end{eqnarray}

Let us now discuss the time evolution of $\hat{U}_{-}(t,t_{0})$.
We introduce a generic projection operator which has the following properties:
\begin{eqnarray}
   P^2 &=& P, \\
   Q &=& 1-P, \\
   PQ &=& QP = 0.
\end{eqnarray}
These are the same as those introduced in \S 2.
%Similarly to Eq. (\ref{eqn:P+Q}),  
By using the projection operators, 
we obtain the following differential equation:
\begin{eqnarray}
\frac{d}{dt}\hat{U}_{-}(t,t_{0}) 
= \hat{U}_{-}(t,t_{0})Pi\breve{L}_{I}(t,t_{0}) 
  + \hat{U}_{-}(t,t_{0})Qi\breve{L}_{I}(t,t_{0}). \label{eqn:usori}
\end{eqnarray}
We can construct the differential equation for $\hat{U}_{-}(t,t_{0})Q$ 
and solve it as we have done in \S 2: 
\begin{eqnarray}
\hat{U}_{-}(t,t_{0})Q 
&=& Q\hat{u}_{-}(t,t_{0})
+\int^{t}_{t_{0}}ds 
\hat{U}_{-}(s,t_{0})Pi\breve{L}_{I}(s,t_{0})Q\hat{u}_{-}(t,s) \label{eqn:ustc} \\
&=& [Q\hat{u}_{-}(t,t_{0}) - \hat{U}_{-}(t,t_{0})P(\hat{\Theta}^{-1}_{-}(t,t_{0})-1)]
\hat{\Theta}_{-}(t,t_{0}),\label{eqn:ustcl}
\end{eqnarray}
where
\begin{eqnarray}
\hat{u}_{-}(t,t_{0}) 
&=& e^{\int^{t}_{t_{0}}ds i\breve{L}_{I}(s,t_{0})Q}_{\rightarrow}, \\
\hat{U}_{+}(t,t_{0}) 
&=& e^{-\int^{t}_{t_{0}}ds i\breve{L}_{I}(s,t_{0})}_{\leftarrow}, \\
\hat{\Theta}_{-}(t,t_{0}) 
&=& [1-\int^{t}_{t_{0}}ds \hat{U}_{+}(t,s)Pi\breve{L}_{I}(s,t_{0})Q\hat{u}_{-}(t,s)]^{-1}.
\end{eqnarray}
When we substitute Eq. (\ref{eqn:ustc}) into (\ref{eqn:usori}), 
we can obtain the TC equation.
On the other hand, substituting Eq. (\ref{eqn:ustcl}) into (\ref{eqn:usori}), 
the TCL equation can be derived.
Here we discuss only the TCL equation.
We obtain
\begin{eqnarray}
\frac{d}{dt}\hat{U}_{-}(t,t_{0}) 
= \hat{U}_{-}(t,t_{0})Pi\breve{L}_{I}(t) 
  - \hat{U}_{-}(t,t_{0})P(1-\hat{\Theta}_{-}(t))i\breve{L}_{I}(t)
  +Q\hat{u}_{-}(t,t_{0})\hat{\Theta}_{-}(t)i\breve{L}_{I}(t). \nonumber \\
\end{eqnarray}
When we substitute this result into Eq. (\ref{eqn:t-bunri}), 
we obtain
\begin{eqnarray}
\frac{d}{dt}O(t)
&=& e^{i\int^{t}_{t_{0}}ds L(s,t_{0})}_{\rightarrow}iL_{0}O(t_{0}) 
+ e^{i\int^{t}_{t_{0}}ds L(s,t_{0})}_{\rightarrow}(U_{0}(t,t_{0}))^{\dagger}Pi\breve{L}_{I}(t,t_{0})(U_{0}^{-1}(t,t_{0}))^{\dagger}O(t_{0}) 
\nonumber \\
&&- e^{i\int^{t}_{t_{0}}ds L(s,t_{0})}_{\rightarrow}(U_{0}(t,t_{0}))^{\dagger}P(1-\hat{\Theta}_{-}(t,t_{0}))i\breve{L}_{I}(t)
(U_{0}^{-1}(t,t_{0}))^{\dagger}O(t_{0}) \nonumber \\
&&+Q\hat{u}_{-}(t,t_{0})\hat{\Theta}_{-}(t,t_{0})i\breve{L}_{I}(t,t_{0})(U_{0}^{-1}(t,t_{0}))^{\dagger}O(t_{0}). \label{eqn:u-s-gen}
\end{eqnarray}
This equation is equivalent to the Heisenberg equation of motion 
and corresponds to Eq. (\ref{eqn:genmitu-nt}).
Up to second order, we have
\begin{eqnarray}
\frac{d}{dt}O(t)
&=& e^{i\int^{t}_{t_{0}}ds L(s,t_{0})}_{\rightarrow}iL_{0}O(t_{0}) 
+ e^{i\int^{t}_{t_{0}}ds L(s,t_{0})}_{\rightarrow}(U_{0}(t,t_{0}))^{\dagger}Pi\breve{L}_{I}(t,t_{0})(U_{0}^{-1}(t,t_{0}))^{\dagger}O(t_{0}) 
\nonumber \\
&&+ e^{i\int^{t}_{t_{0}}ds L(s,t_{0})}_{\rightarrow}(U_{0}(t,t_{0}))^{\dagger}P\int^{t}_{t_{0}}ds 
i\breve{L}_{I}(s,t_{0})Qi\breve{L}_{I}(t)
(U_{0}^{-1}(t,t_{0}))^{\dagger}O(t_{0}) \nonumber \\
&&+Q\hat{u}_{-}(t,t_{0})\hat{\Theta}_{-}(t,t_{0})i\breve{L}_{I}(t,t_{0})(U_{0}^{-1}(t,t_{0}))^{\dagger}O(t_{0}). \label{eqn:u-s-app}
\end{eqnarray}
Here, we do not expand the third term on the r.h.s.~of Eq. (\ref{eqn:u-s-app}).

The derivation given here is not the exactly same as that 
proposed by Uchiyama and Shibata.
We have modified the following two points.
In the original paper of Uchiyama and Shibata, 
they imposed the condition
\begin{eqnarray}
(U_{0}^{-1}(t,t_{0}))^{\dagger}O(t_{0}) = f(t,t_{0})O(t_{0}),
\end{eqnarray}
where $f(t,t_{0})$ is a c-number function, and 
$O(t_{0})$ is the operator seen on the r.h.s.~of Eq. (\ref{eqn:u-s-gen}).
However, we have not used this condition in the derivation given here.
Furthermore, they imposed the condition
\begin{eqnarray}
PO(t_{0})=O(t_{0}).
\end{eqnarray}

It is easily seen that no condition needs to be imposed to derive the extended equation in the extended U-S method, 
while in the method given in this paper 
the condition $QL_{0}Q=QL_{0}$ is required.
However, there is a merit to our method: 
All quantities which commute with the total Hamiltonian 
are conserved in the time evolution regardless of the order of 
the expansion of the equation.
It is not clear that this property is satisfied in the extended U-S method.

This can be shown concretely.
We consider the following Hamiltonian:
\begin{eqnarray}
H = \omega_{a}a^{\dagger}a + \omega_{b}b^{\dagger}b 
 +g( a^2b^{\dagger 2} + a^{\dagger 2}b^2 ).
\end{eqnarray}
The operator $a$ is the degree of freedom of the system, and 
the operator $b$ is that of the environment.
The total Hamiltonian can be divided into the following three parts:
\begin{eqnarray}
H_{S} &=& \omega_{a}a^{\dagger}a, \\
H_{E} &=& \omega_{b}b^{\dagger}b, \\
H_{I} &=& g( a^2b^{\dagger 2} + a^{\dagger 2}b^2 ).
\end{eqnarray}
We use the same projection operator as in Eq.~(\ref{eqn:ex-pro}).
Now, we calculate the time evolution of the total Hamiltonian $H$ 
in the extended U-S method.
We expand $(1-\hat{\Theta}_{-}(t))$ to zeroth order in the interaction 
in Eq. (\ref{eqn:u-s-gen}).
This means that we ignore the third term on the r.h.s.~of the equation.
The equation of motion is expressed as 
\begin{eqnarray}
\frac{d}{dt}\langle H (t)\rangle 
&=& \langle e^{iLt}iL_{0}H +e^{iLt}e^{-iL_{0}t}Pe^{iL_{0}t}iL_{I}H \rangle \nonumber \\
&=& 2ig(\omega_{a}-\omega_{b})
   (-\langle a^2 (t)b^{\dagger 2}(t) \rangle 
    +\langle a^{\dagger 2}(t)b^2(t) \rangle  \nonumber \\
&&    + e^{2i\omega_{b}t}\langle a^2(t) \rangle\langle b^{\dagger 2} \rangle
    - e^{-2i\omega_{b}t}\langle a^{\dagger 2}(t)\rangle \langle b^2 \rangle ), \nonumber \\
\end{eqnarray}
where we take $t_{0}=0$.
The total Hamiltonian has no time dependence originally.
However, 
it is not obvious whether the r.h.s.~of the equation becomes zero for 
arbitrarily states.

On the other hand, 
the time dependence of the Hamiltonian calculated using our method 
vanishes: $\frac{d}{dt}H(t) = 0$.
This can be easily shown.
To begin, we substitute $O=H$ into Eq.~(\ref{eqn:t-i-res}).
Then, we must calculate the commutation relation $LO=LH$ first.
This quantity clearly becomes zero.
This is satisfied regardless of the order of 
the expansion and Hamiltonian considered.
Similarly, other quantities that commute with the total Hamiltonian 
are also conserved.

\end{document}